\documentclass[a4paper,12pt]{article}

\usepackage{graphicx}
\usepackage{amsmath,amssymb}
\usepackage{a4wide}
\usepackage{cite}
\usepackage{subfigure}

\begin{document}

\begin{center}

   {\large \bf Stress-driven oxidation chemistry of wet silicon surfaces}

   \vspace{0.5cm}
    Lucio Colombi Ciacchi,$^{1,2,*}$
    Daniel J. Cole,$^{3}$ 
    Mike C. Payne,$^{3}$
    Peter Gumbsch$^{1,2}$

   \vspace{0.25cm}
   {\it
       $^1$Fraunhofer Institut f\"ur Werkstoffmechanik, 
           W\"ohlerstrasse 11, 79108 Freiburg, Germany \\
       $^2$Institut f\"ur Zuverl\"assigkeit von Bauteilen und Systemen,
           Universit\"at Karlsruhe, Kaiserstr. 12, 76131 Karlsruhe, Germany. \\ 
       $^3$Theory of Condensed Matter Group, Cavendish Laboratory,
             J J Thomson Avenue,  Cambridge CB3 0HE, UK}

\vspace{0.5cm} \today 

\end{center}

\begin{center}
  {\bf Abstract}
\end{center}

The formation of a hydroxylated native oxide layer on Si(001) under wet 
conditions is studied by means of first principles molecular dynamics simulations.
Water molecules are found to adsorb and dissociate on the oxidised surface leading 
to rupture of Si-O bonds and producing reactive sites for attack
by dissolved dioxygen or hydrogen peroxide molecules.
Tensile strain is found to enhance the driving force for the dissociative adsorption
of water, suggesting that similar reactions could be responsible for environmentally-driven 
sub-critical crack propagation in silicon.

\vfill \noindent
$^*$ To whom correspondence should be addressed: colombi@hmi.uni-bremen.de

\clearpage

The chemistry at the interface between silicon and a wet environment 
is of great importance for the processing, the function and 
the failure of microelectromechanical systems (MEMS).
In particular, the behaviour of MEMS for biomedical applications~\cite{Grayson_04}
is governed by the adhesion of extracellular matrix proteins 
to the oxidised surface~\cite{Davis_02} and by dissolution of
Si(OH)$_{4}$ molecules from the surface~\cite{Anderson_03}.
Similar dissolution processes occur during the release of MEMS via etching of 
sacrificial oxides.
During this processing stage, dissolved  oxygen in the etching solution 
may lead  to pronounced surface roughness~\cite{Wade_97,Garcia_02}, which may 
later promote crack initiation and failure of the device under mechanical load~\cite{Chen_02}.
Moreover, when mobile devices operate in a humid atmosphere, the crack-propagation rate
under fatigue loading is substantially accelerated and the device's life-time
reduced with respect to vacuum conditions~\cite{Bagdahn_01,Muhlstein_02,Allameh_03,Kahn_02}.

Conclusive atomic-level interpretations for the origin of all these 
processes are lacking, due to the complexity of phenomena which include a combination 
of surface oxidation, reaction of the native oxide layer with water molecules,
and the presence of mechanical stresses in the silicon structure~\cite{Muller_00}.
To gain an atomistic insight into these mechanisms we have studied the oxidation of a 
Si(001) surface, the reactions of the oxidised surface with liquid water and dissolved 
oxidising chemical species, as well as the effect of mechanical  stresses on the reactivity 
of the surface by means of  first-principles molecular dynamics (FPMD) 
simulations~\cite{Car_85,Payne_92,DFT_details}.

Under dry conditions, we have found that oxygen molecules spontaneously react with the 
bare surface and a native oxide layer quickly grows until the oxygen coverage reaches 
about 1.5~monolayer (ML)~\cite{Colombi_05}.
At higher coverages, further oxide growth is limited by the thermally activated diffusion 
of dioxygen molecules through the oxide layer to the chemically active Si/SiO$_x$ 
interface~\cite{Bongiorno_04}.
Namely, several attempts at placing O$_{2}$ molecules close to putative adsorption
sites all resulted in repulsion of the added molecule away from the oxide layer.
Moreover, no changes in the topology of the Si-O network have been observed
upon annealing of the structure at 600~K in a 0.6~ps FPMD simulation.
These results indicate a considerable chemical stability of the oxide model,
which has a Si:O ratio of about 1:1 and contains partially oxidised 
Si species in proportions roughly consistent with medium-energy ion scattering 
experiments~\cite{Hoshino_01} (details on the composition obtained are
reported in Refs.~\cite{Colombi_05} and~\cite{Cole_07}).

Oxide layers of very similar structure and composition form also when B and P 
impurities are initially present in the bare surface layers.
In this case the oxidation proceeds while the impurities remain
trapped at the Si/SiO$_x$ interface and an oxide layer gradually 
builds up above them~\cite{Cole_07}.
This indicates that dopants should not substantially influence the interactions
between the oxide layer and an outer wet environment and are thus not considered
in the simulations presented here.

Interestingly, the formation of a thin oxide layer on Si(001) is accompanied by 
development of surface tensile stress~\cite{Cole_07}, as previously 
measured experimentally~\cite{Sander_91}.
In our simulations, the surface stress initially decreases from the 
value of 0.81~N/m for the bare p-2x2 reconstructed Si(001) surface 
to a minimum of 0.05~N/m at an oxygen coverage of 0.75~ML, 
then it rapidly increases at higher coverages, reaching a maximum 
of 2.92~N/m at 1.5~ML.
The fact that a tensile stress is present in the amorphous SiO$_{x}$ layer which
spontaneously forms on the silicon surface in contact with the atmosphere 
suggests that the surface could readily react with water molecules, as in the case 
of mechanically stressed silicate glasses~\cite{Gy_03}.
In silicate glasses, attack by water is thought to be due to polarisation of a 
strained  Si-O bond, which promotes the adsorption of water on the Si atom accompanied
by proton transfer to a neighbouring O atom, eventually resulting in the
cleavage of the bond~\cite{Michalske_82}.

To test whether similar reactions may take place on the native oxide grown 
on Si surfaces, we have performed a FPMD simulation of the model
previously obtained for the dry oxidised Si(001) surface in contact with 
liquid water (Fig.~1).
The atomic coordinates of the oxidised models were directly taken from the 
output of the calculations described in Ref.~\cite{Colombi_05},
while the positions of the water molecules were initially randomised.
As noted above, in spite of the presence of exposed Si$^{2+}$ and Si$^{3+}$ 
species, the oxidised surface did not present any reactive sites for further reactions 
with  O$_2$ molecules~\cite{Colombi_05}.
However, in the present case, soon after starting the simulation we observe
two water molecules adsorbing on the surface. 
The first molecule binds with its O atom to a twofold-coordinated Si atom
donating a proton to a neighbouring water molecule.
This proton then stays in solution for the remainder of the simulation,
indicating an active role of the water hydrogen-bond network in the
reaction mechanism~\cite{Ma_2005}.
The second molecule binds to a fourfold-coordinated Si atom, initially 
forming a fivefold-coordinated reaction intermediate (Fig.~1a).
This adsorbed molecule later dissociates donating a proton to a nearby
oxygen atom, which was originally bridging two Si atoms of the oxide 
structure.
As a consequence, one of the Si-O bonds breaks with formation of a 
hydroxyl group bound to a Si atom on one side and of a 
threefold-coordinated Si atom on the other side (Fig.~1b).

The creation of this newly exposed reactive site opens up the possibility 
of further reactions with water molecules or other chemical species which 
may be dissolved in solution, such as hydrogen peroxide or oxygen.
Indeed, continuing the simulation, a third water molecule spontaneously 
adsorbs on this site (Fig.~1c) and in turn dissociates donating a 
proton to the Si atom which was originally twofold-coordinated in the dry 
oxide model (Fig.~1d).
Moreover, in further FPMD simulations we replaced this third water molecule
with an H$_2$O$_2$ molecule prior to the adsorption event (with an initial
Si-O distance  of 2.8~\AA\ between the molecule and the newly created
threefold-coordinated Si site).
As for the case of H$_{2}$O, the H$_2$O$_2$ molecule quickly binds to the 
surface and dissociates (Fig.~2a-d).
Unexpectedly, not only does cleavage of the O-O bond of H$_2$O$_2$ occur upon
chemisorption but also proton transfer from the dissociating molecule to the
originally twofold-coordinated Si atom, just as was observed in the case of water 
adsorption (cf. Fig.~2b with Fig.~1d).
This suggests that concerted electrophilic/nucleophilic attacks could also take
place during the stress corrosion of silicate glass induced by H$_{2}$O$_{2}$,
and provides an alternative mechanism to the proposed simple cleavage of the 
peroxide molecule leading to the creation of two OH radicals~\cite{Guyer_03}.

Finally, when an O$_{2}$ molecule instead of an H$_2$O$_2$ molecule is
placed near the same reactive site, spontaneous chemisorption and dissociation
again take place (Fig.~2e-h).
During the reaction, which is driven by the donation of electrons from the surface
to the antibonding molecular orbitals of O$_2$~\cite{Colombi_05}, a proton is 
donated to the dissociating molecule from a hydroxyl group nearby (Fig.~2f).
Later, the O$^-$ atom produced receives, from the solution, a dissolved proton 
which had been released previously by a water molecule after adsorption on 
the twofold-coordinated Si surface site (Fig.~2f,h).
After the dissociation, the second O atom binds to the originally twofold-coordinated 
Si atom, revealing in particular the ability of the latter to be both reduced (i.e. to 
accept protons) and oxidised (cf. Fig.~2d and Fig.~2h).

At the end of the simulation the hydroxylated oxide layer is composed of 13 oxidised 
Si species with formal valences ranging from +1 to +4, 12 O atoms twofold coordinated 
by Si atoms, and 4 adsorbed OH groups.
It is worth noting that, despite the limited size of the model and the short simulation 
time, the obtained concentration of 4 hydroxyl groups on a surface of 1.2~nm$^{2}$
agrees well with the values between 2.6 and 4.6 OH/nm$^{2}$ measured experimentally 
on amorphous silica surfaces~\cite{Souza_99,Zhuravlev_00}.

In order to check whether the presence of stress in the native oxide structure
has a direct influence on the dissociative adsorption of water and the creation of 
exposed reactive Si sites, we have repeated the simulation shown in Fig.~1 in 
the presence of -7.5~\%\ compressive plane strain.
Strain was applied via a uniform rescaling of the cell vectors and of the atomic 
coordinates of the oxidised Si model in the directions parallel to the surface.
After relaxing the atomic positions, the height of the simulation cell was adjusted 
to obtain a free volume equal to the case of zero-strain, which was filled with the
same number of water molecules as before. 
In this case, two water molecules are again observed to adsorb on Si atoms of
the surface, one of which donates a proton into the bulk water.
However, whereas in the absence of external strain (Fig.~1) protons were 
transferred to O atoms of the surface and Si-O bonds broken, in this case neither 
processes are observed during the entire simulation, which was stopped after 3.3~ps.
In addition, we performed a further FPMD simulation starting from the final structure
obtained in Fig.~1(d) and applying an external tensile plane strain of 7.5~\%~\cite{stress}.
In this case, after about 0.6~ps, a third water molecule was observed to attack 
a strained Si-Si bond, adsorbing onto one of the two atoms, dissociating and donating 
a proton to the second atom, irreversibly breaking the bond.
These combined results suggest that the driving force for water attack of 
surface bonds may be reduced by the presence of compressive 
and enhanced by the presence of tensile externally applied stress.

We would like to note here that the simulations performed above are only
indicative of the possible chemical processes that occur, and cannot provide
any quantitative conclusions regarding reaction times and reaction rates.
Moreover, in all cases we started from randomized positions of water molecules,
which may in principle influence the chemical mechanisms that are observed here.
Nevertheless, it is interesting that, both in the simulation in the presence 
of compressive strain and in the simulation in the absence of strain, a water 
molecule adsorbs on the same Si surface site, although at a later simulation 
time when compression is present (about 1.0 ps vs. 0.3 ps, respectively).
However, only in the absence of compressive strain does the molecule dissociate causing
the  rupture of a Si-O bond, suggesting a strong influence of the strain field
on the behavior of the system.

To substantiate this hypothesis we have performed FPMD simulations of a system 
composed of the oxidised surface and a single water molecule initially placed at a 
distance of 2.8~\AA\ from the fourfold-coordinated Si atom which first reacted with 
water in the simulation described above (labelled 5 in Figure 1a).
In the absence of externally applied strain or in the presence of tensile strain, the water
molecule adsorbs spontaneously immediately after starting the simulation.
Remarkably, when the surface slab is put under compressive strain, no spontaneous 
adsorption takes place and the molecule diffuses away from the surface.

To understand the reasons behind this observation, we have computed the atomic
point charges which best fit the electrostatic potential obtained {\em ab initio} in a 
1~\AA\ thick region outside the Van der Waals radius of the atoms of the oxidised 
surface (ESP charges~\cite{Cox_1984}).
The positive ESP charge on the Si atom is found to {\em increase} from a value
of 0.06~$e$ under compressive strain to 0.13~$e$ in the absence of strain and to 0.20~$e$
in the case of tensile strain, indicating progressively increasing electrophilicity of the Si site.
Consistently, the charges on neighbouring O atoms are more negative 
by about 0.1~$e$, enhancing the electrostatic driving force toward proton acception.
Analysis of the local density of states (LDOS) projected on the Si atom of the 
surface and on the O atom of the water molecule immediately after formation of the
penta-coordinated complex (see Fig. 1a) shows that the Si-O chemical bond 
initially forms via hybridisation of the p orbitals of O with d-type orbitals of Si
(Fig.~3a).
The associated d-type LDOS peak moves to higher energy and increases in intensity 
when the externally applied strain moves from compression to tension (Fig.~3b),
indicating increased chemical reactivity of the Si atom with increasing
tensile strain. 
In summary,  the presence of tensile stress has the effect of both enhancing the
electrostatic driving force  for the approach of water molecules to Si sites and 
facilitating the hybridisation of Si and O orbitals.
This leads to water chemisorption, subsequent deprotonation and breaking of 
existing Si-O bonds, thus exposing reactive Si sites to the solution environment.

The results of our simulations suggest that further oxidation reactions 
of previously oxidised Si surfaces which are inert under dry conditions 
can take place in the combined presence of tensile stress and humid air.
The enhanced oxidation reactivity of silicon in the presence of a humid 
environment is consistent with the fact that even the very low concentrations 
of dissolved oxygen molecules present in conventional etching solutions can 
significantly affect the morphology of the resulting surfaces, in particular 
leading to pronounced surface roughness~\cite{Wade_97,Garcia_02}. 

Our results seem to support a mechanism for the observed premature failure of 
single-crystal silicon MEMS under fatigue loading in humid environments based 
on sequential steps of oxide formation and stress-driven corrosion cracking of 
the oxide layer at the tip of a stable crack~%
\cite{Muhlstein_02,Allameh_03,Alsem_05,Pierron_06}.
Historically, this ``reaction-layer'' mechanism had been questioned since it was 
thought that a relatively thick ($\sim$20~nm) oxide layer was necessary to 
activate stress-corrosion, and the time-scale of the fatigue process is too short 
to allow such thick oxide growth~\cite{Kahn_02,Kahn_04}.
However, as we have now demonstrated, ultrathin oxide layers on the Si(001) surface
are under tensile stress at coverages of about 1~ML of oxygen, and are subjected
to virtually barrierless stress-driven water attack leading to breaking 
of Si-O bonds.
Whether or not such mechanisms are active in silicon MEMS may depend
on the crystallographic directions of the propagating crack.
Indeed, while both our calculations and experimental evidence~\cite{Sander_91} 
show that tensile stress is naturally present on the oxidised (001) surface, only
compressive stress was found to develop at all oxygen coverages upon oxidation 
of the (111) surface  at room temperature ~\cite{Sander_91}.
Finally, we note that more complex mechanisms involving grain boundaries may be 
active during sub-critical crack propagation in polycrystalline silicon structures, where
fatigue is found to occur irrespective of the presence of a corrosive environment~%
\cite{Bagdahn_03,Kahn_06}.

\section*{Acknowledgements}
Computer time was provided by the HLRS, Stuttgart, and by the ZIH, Dresden, Germany.
This work has been partially supported by the HPC-EUROPA project.
LCC acknowledges funding from the Alexander von Humboldt Foundation and
the Deutschen Forschungsgemeinschaft within the Emmy-Noether Programme.

\clearpage

\section*{Figures}

\begin{figure}[h!]
 \caption{Snapshots from a FPMD simulation of a Si(001) 
surface covered by its native oxide layer in contact with water. 
Si atoms are grey, O red,  H white.
(a) adsorption  of a water molecule on a Si atom (labelled 5) forming a 
penta-coordinate reaction intermediate (simulation time $t=0.5$~ps).
(b) dissociation of the water molecule (see arrow) resulting in breaking of 
a Si-O bond nearby, formation of two Si-OH groups, and exposure of a 
threefold-coordinated Si atom to the solution medium (labelled 3) ($t=1.1$~ps).
(c) adsorption of a further water molecule (black arrow) on the newly exposed
reactive site ($t=1.4$~ps).
(d) dissociation of the latter water molecule causing protonation (black arrow) 
of a Si atom nearby (labelled 2) ($t=2.9$~ps). }
\end{figure}

\begin{figure}[h!]
 \caption{Snapshots from two FPMD simulations of the reaction 
of  H$_2$O$_2$ (a-d) and  O$_2$ (e-h) with an electrophilic Si site of the oxidised 
Si(001) surface (labelled 3) exposed upon chemical attack of a Si-O bond by water 
(see Fig.~1a,b).
Color code and labels are as in Fig.~1. The O atoms of the reacting molecules
are indicated with asterisks, and proton transfer processes are indicated with
black arrows (see text).}
\end{figure}

\begin{figure}[h!] 
 \caption{(a) Local density of states (LDOS) projected on the d 
 orbitals of a surface Si atom and on the p orbitals of the O atom of a water
 molecule chemisorbing on it.
(b) Shift to higher energy and increase of intensity of the d-type LDOS peak
of the same Si atom in the presence of applied strain.}
\end{figure}

\clearpage

\begin{center}
 \includegraphics[width=\textwidth]{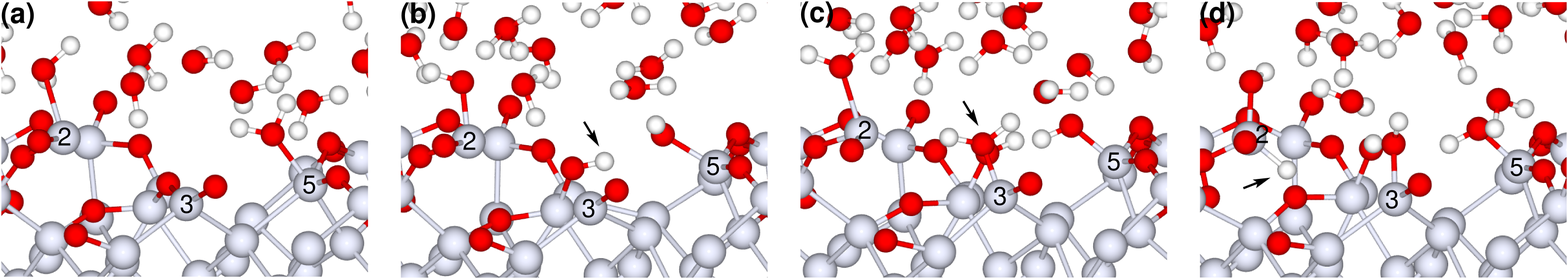}
\vfill\noindent
L. Colombi Ciacchi et al., Figure 1
 \end{center}
 
\clearpage

\begin{center}
 \includegraphics[width=\textwidth]{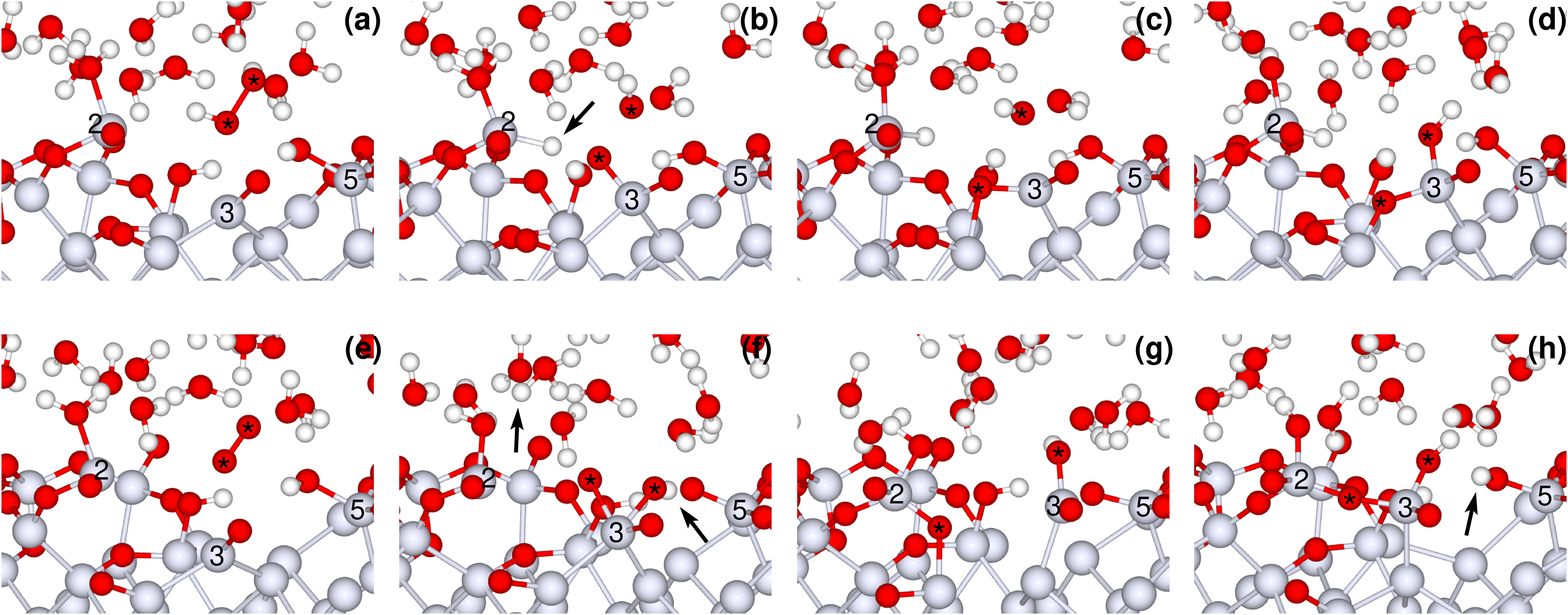}
 \vfill\noindent
L. Colombi Ciacchi et al., Figure 2
 \end{center}
 
 \clearpage
 
 \begin{center}
\includegraphics[width=0.5\textwidth]{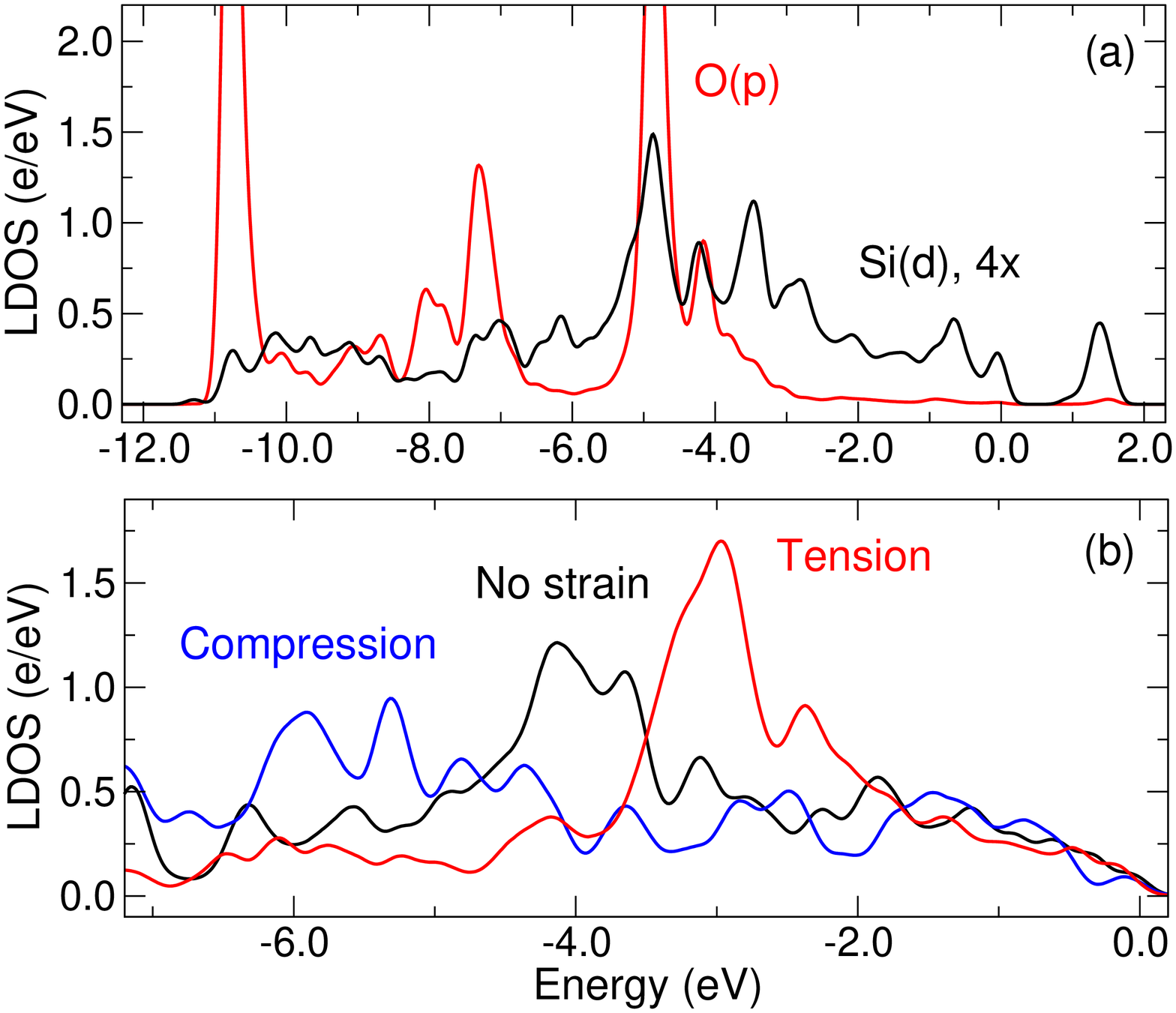}
\vfill\noindent
L. Colombi Ciacchi et al., Figure 3
\end{center}

\clearpage
 
 \begin{center}
\includegraphics[width=0.4\textwidth]{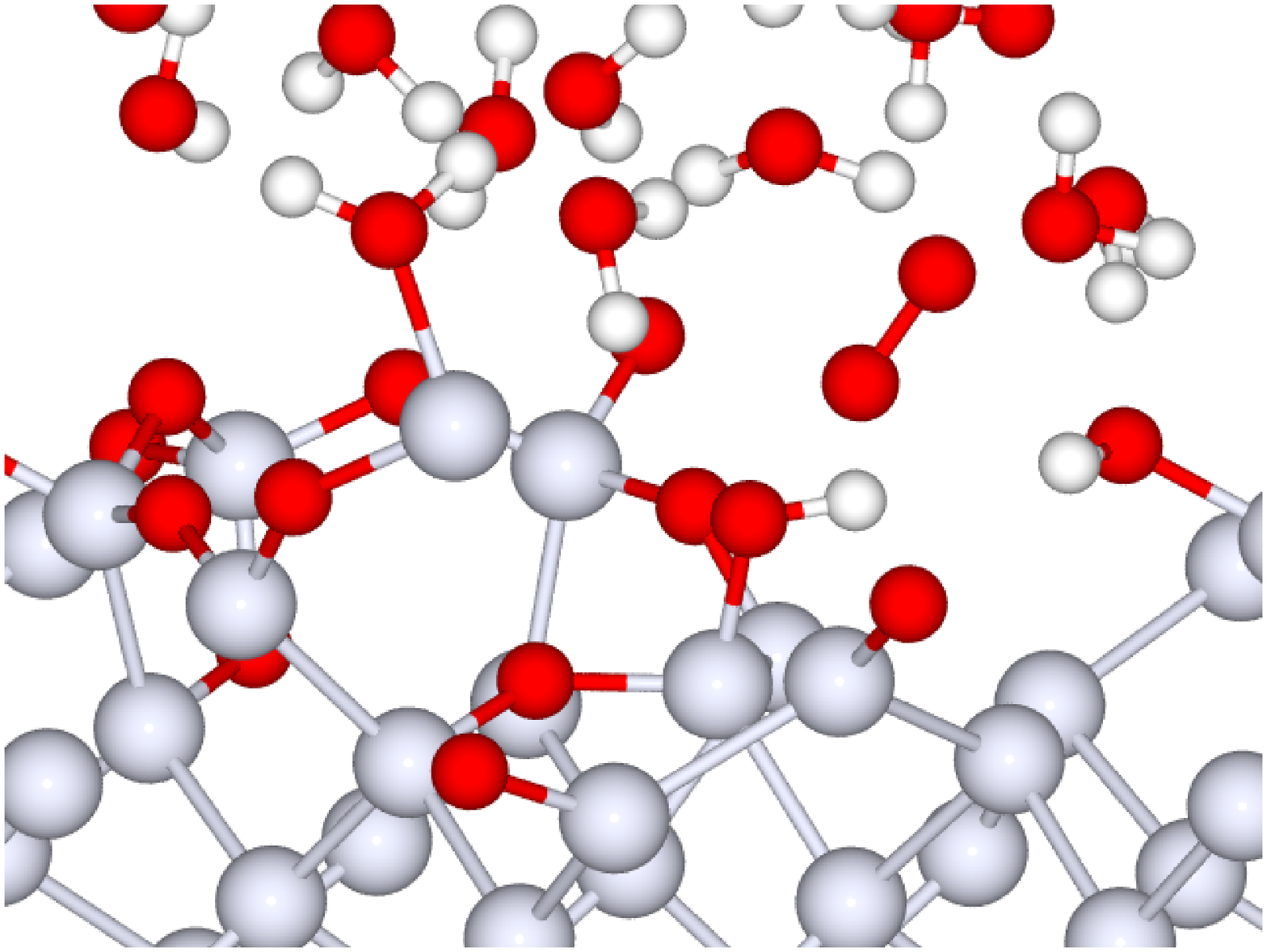}
\vfill\noindent
L. Colombi Ciacchi et al., Table of Contents Image
\end{center}


\begin{thebibliography}{99}
\bibitem{DFT_details}
Our density-functional theory simulations were performed using 
norm-conserving pseudopotentials and the PW91 xc potential.
%
The wave functions were expanded in plane waves at the (0.25,0.25,0.0) point
of the Brillouin zone up to a kinetic energy cut-off of 60~Ry, applying a
smearing of $k_{B}T\!=\!0.1$~eV to the electronic occupancies.
%
A convergence test of the forces and the MD trajectory during the adsorption
of a water molecule on the oxidised Si(100) surface has been performed using
a larger cut-off of 80~Ry.
%
Differences between forces are limited to a few meV/\AA\ and differences between
atom-atom distances to about 0.01~\AA\, indicating that 60~Ry is an acceptable
cut-off for the simulation of the adsorption reactions investigated here.
%
All dynamical simulations were performed with the Car-Parrinello method~\cite{Car_85} 
using a time-step of 5.0 au, a fictitious electronic orbital mass of 300 au
and a hydrogen mass of 2.0~amu.
%
Using a deuterium mass for H atoms is usual in Car-Parrinello simulations
including water, and ensures that adiabatic conditions for the electrons
are maintained throughout the dynamics~\cite{Silvestrelli_99}.
%
The Si(001) surface was modelled with a periodically repeated
slab of 8 Si layers separated by 8 corresponding layers of vacuum,
using a 2x2 surface unit cell with surface area of 10.91x10.91~\AA$^{2}$.
%
In the simulations including water the space between the slabs was filled 
with 29 water molecules (corresponding to a density of about 1 g/cm$^{3}$)
and the bottom surface was passivated with H atoms.
%
All simulations were performed without temperature constraint, the 
simulation temperature never exceeding 350~K.
\end{thebibliography}
\end{document}